\documentclass[preprint,aps,prl,preprintnumbers,amsmath,amssymb,superscriptaddress,showpacs,floatfix]{revtex4}
\usepackage{epsfig}
\usepackage{dcolumn}
\usepackage{bm}
\usepackage{amssymb}
\usepackage[bookmarks=false]{hyperref}

\begin{document}

\title{Connecting Polyakov Loops to the Thermodynamics of $SU(N_c)$ Gauge Theories Using the Gauge-String Duality}

\author{Jorge Noronha}
\affiliation{Department of
Physics, Columbia University, 538 West 120$^{th}$ Street, New York,
NY 10027, USA}

\begin{abstract}
We show that in four-dimensional gauge theories dual to five-dimensional Einstein gravity coupled to a single scalar field in the bulk the derivative of the single heavy quark free energy in the deconfined phase is $dF_{Q}(T)/dT \sim -1/c_s^2(T)$, where $c_s(T)$ is the speed of sound. This general result provides a direct link between the softest point in the equation of state of strongly-coupled plasmas and the deconfinement phase transition described by the expectation value of the Polyakov loop. We give an explicit example of a gravity dual with black hole solutions that can reproduce the lattice results for the expectation value of the Polyakov loop and the thermodynamics of $SU(3)$ Yang-Mills theory in the (non-perturbative) temperature range between $T_c$ and $3T_c$.

\end{abstract}

\pacs{12.38.Mh, 11.25.Tq, 11.25.Wx, 24.85.+p}
\maketitle


The conjectured duality between gauge fields and strings may hold the key to understand the complicated dynamics of non-Abelian gauge theories \cite{Polyakov:1997tj}. The Anti-de Sitter/Conformal Field Theory (AdS/CFT) correspondence \cite{Maldacena:1997re,Aharony:1999ti} provides an explicit example where one of the main properties of $\mathcal{N}=4$ Supersymmetric Yang-Mills (SYM) in four dimensions, i.e., its exact conformal invariance, is already enough to determine that the background spacetime where the dual strings propagate contains a 5-dimensional AdS space. In this case, the dual string theory is a (critical, i.e., 10 dimensional) type IIB string defined in AdS$_5\otimes S_5$. Unfortunately, QCD and $\mathcal{N}=4$ SYM have very different properties even when $N_c\to \infty$. For instance, while the former exhibits confinement below $T_c$ and asymptotic freedom at high temperatures, the latter is always deconfined at any temperature in an infinite volume \cite{Witten:1998zw}. Moreover, in the region between $T_c$ and $4T_c$ lattice calculations have shown that violations from conformal invariance in QCD \cite{Bazavov:2009zn}, or pure Yang-Mills (YM) with $N_c\geq 3$ \cite{largeNclattice}, are maximal. Thus, any realistic string theory dual description of QCD in the temperature range where strong-coupling effects are important must incorporate this important feature observed on the lattice. It remains unknown which type of critical string theory can describe the required infrared and ultraviolet properties of QCD, although some interesting top-down constructions have been proposed in the last years \cite{topdown}. On the other hand, bottom-up phenomenological approaches have also been investigated in detail in the last years \cite{bottomup,andreev}.

Significant progress can be made by considering 5-dimensional gravity theories in which the solutions of the equations of motion describe a spacetime that is asymptotically AdS$_5$, which in turn implies conformal invariance in the UV for the gauge theory at the boundary. The minimal realization of this idea corresponds to Einstein gravity coupled to a nontrivial scalar field, $\phi$, which is used to describe the violation of conformal invariance in the infrared \cite{Csaki:2006ji,Gubser:2008ny,pufu,Gursoy:2007cb,Gursoy:2008bu,Hohler:2009tv,Cherman:2009kf}. In fact, it was shown in \cite{Gubser:2008ny} that there is a class of scalar potentials $V(\phi)$ that generate black hole solutions with thermodynamic properties very similar to finite temperature QCD lattice results. According to the AdS/CFT duality dictionary, the non-Abelian 4-dimensional theories dual to these gravity theories are CFT's modified by the addition of a term $\sim \Lambda^{4-\Delta}\mathcal{O}_{\phi}$, where $\Lambda$ is the energy scale of the deformation and $\Delta < 4$ is the UV scaling dimension of the relevant operator $\mathcal{O}_{\phi}$ dual to $\phi$. One may interpret this gravitational model as a type of holographic ``effective theory" for QCD valid at energy scales where effects from asymptotic freedom, which may not be fully described in the supergravity approximation \cite{Aharony:1999ti}, are negligible. Alternatively, such models appear quite naturally as possible realizations of non-critical string theories in 5 dimensions as discussed in detail in \cite{Csaki:2006ji,Gursoy:2007cb}. In a sense, currently gravity duals and perturbative QCD (pQCD) provide complementary descriptions of a plethora of QCD phenomena. 

Even though $V(\phi)$ is not known from first principles, its form can be somewhat constrained by requiring the dual gauge theory to display some important features observed in YM theories. In fact, there is a simple argument that justifies looking for potentials that have a simple exponential form $V\sim - e^{b \phi}$ \cite{Csaki:2006ji}, where $b$ is a positive number. The idea is to assume that the gravity dual's action comes from a non-critical string theory in $D=5$ dimensions and, thus, there should be a constant term that goes like $\alpha'\,(10-D)$ in the string frame action. This means that the Einstein frame potential $V$ may have (at least) a term that goes like $e^{\sqrt{2/3}\,\phi}$. In fact, the potentials studied in \cite{Gubser:2008ny} are such that $V(\phi)\sim -e^{b \phi}$ in the IR and it was shown in \cite{Gursoy:2007cb} that confinement occurs when (among other possible choices) $V(\phi) \sim -e^{\sqrt{\frac{2}{3}}\,\phi}$ for large $\phi$.
 
Here we consider the model used by Gubser and Nellore in \cite{Gubser:2008ny} and use the method (and conventions) they developed to solve the coupled nonlinear equations of motion for the metric, $G_{\mu\nu}$, and the scalar obtained from the action (in the Einstein frame)
\begin{equation}
\ S=\frac{1}{2 k_5^2} \int d^5x\sqrt{-G}\left[\mathcal{R}-\frac{(\partial \phi)^2}{2}-V(\phi)\right].
\label{dilatonaction}
\end{equation}  

Conformal invariance in the UV can be obtained when $\phi\to 0$ at the boundary ($u\to 0$) and
\begin{equation}  
\lim_{\phi\to 0} V(\phi)=-\frac{12}{R^2}+\frac{1}{2 R^2}\Delta(\Delta-4)\phi^2+\mathcal{O}(\phi^4),
\label{conditionboudnary}
\end{equation}
where the mass squared of the scalar $m_{\phi}^2 R^2=\Delta(\Delta-4)\geq -4$ according to the Breitenlohner-Freedman (BF) bound \cite{BFbound}. Eq.\ (\ref{conditionboudnary}) implies that the spacetime will be asymptotically AdS$_5$ with radius $R$. 
        
In this effective theory, the gravitational constant $k_5^2$ is a free parameter expected to be $\sim \mathcal{O}(1/N_c^2)$. The most general ansatz for the line element compatible with the required symmetries is \cite{Gubser:2008ny}
\begin{equation}
\ ds^2 = e^{2A(u)}(-f(u) dt^2+d\vec{x}^2)+e^{2B(u)}\frac{du^2}{f(u)}\,,\qquad \phi=\phi(u).
\label{metric}
\end{equation} 
(note that $e^B$ has dimensions of length). The boundary of the bulk spacetime is located at $u=0$. Finite temperature effects are included by considering solutions that display a horizon at $u=u_h$ where $f(u_h)=0$. The functions above are assumed to be finite at $u_h$. Using this ansatz for the metric the entropy density is given by
\begin{equation}
s=2\pi\,\frac{e^{3A(u_h)}}{k_5^2} 
\label{bekenstein}
\end{equation}
while the temperature is (we use a prime for $d/du$)
\begin{equation}
T=\frac{e^{A(u_h)-B(u_h)}|f'(u_h)|}{4\pi} \,.
\label{hawking}
\end{equation}

We follow the gauge choice used in \cite{Gubser:2008ny} where $\phi=u$. The equations of motion in this gauge are
\begin{eqnarray}
A''-A'B'+\frac{1}{6} & = & 0\,, \label{EOM1}\\
f''+(4A'-B')f' & = & 0\,,\label{EOM2}\\
6A' f'  + f (24A'^2-1)+2e^{2B}V & = & 0\,, \label{EOM3}\\
4A'-B'+\frac{f'}{f}-\frac{2e^{2B}}{f}V' & = & 0\,.\label{EOM4}
\end{eqnarray} 
Gubser and Nellore showed that these equations can be solved using a generating function defined as $G(u)=A'(u)$. In fact, 
\begin{equation}
\ A(u)=A_h + \int_{u_h}^{u}d\xi \,G(\xi) 
\end{equation}
and, from (\ref{EOM1}) we obtain
\begin{equation}
\ B(u)=B_h + \int_{u_h}^{u}d\xi \,\frac{G'(\xi)+1/6}{G(\xi)} 
\end{equation}
and $f(u)$ can be found from (\ref{EOM2}) and it reads
\begin{equation}
\ f(u)=f_1\int_{u_h}^{u}d\xi \,e^{-4A(\xi)+B(\xi)} 
\end{equation}
where $A_h$, $B_h$, $f_1$ do not depend on $u$. Eq.\ (\ref{EOM3}) leads to an expression for $V$ in terms of the quantities derived above
\begin{equation}
\ V(u)=f(u)\frac{e^{-2B(u)}}{2}\left(1-24 G(u)^2 -6 G(u)\frac{f'(u)}{f(u)}   \right).
\end{equation}
The last equation of motion (\ref{EOM4}) does not add any new information at this point.

This method to solve Einstein's equations becomes can be very useful because it is possible to show \cite{Gubser:2008ny} that the equations above lead to the following nonlinear differential equation for $G$ in terms of $V$
\begin{equation}
\frac{G'}{G+\frac{V}{3V'}}=\frac{d}{du}\log \left(\frac{G'}{G}+\frac{1}{6G}-4G-\frac{G'}{G+\frac{V}{3V'}}\right).
\label{nonlinear}
\end{equation} 
Thus, the full solution of Einstein's equations for a given $V$ can be directly obtained by solving the equation above for $G$. This can be done by choosing a value for $u_h$ and then seeding a numerical integrator subroutine using a series solution of (\ref{nonlinear}) around $u_h$. Such an expansion can be found using that Eqs.\ (\ref{EOM3}) and (\ref{EOM4}) when evaluated at the horizon become
\begin{eqnarray}
\ V(u_h)&=&-3 e^{-2B(u_h)}G(u_h)f'(u_h)\\
V'(u_h)&=&e^{-2B(u_h)}f'(u_h)\,,
\label{constraints}
\end{eqnarray}
which lead to the following identity
\begin{equation}
G(u_h)=-\frac{V(u_h)}{3V'(u_h)}
\label{constraint1}
\end{equation}   
and the desired series
\begin{eqnarray}
G(u)&=& -\frac{1}{3}\frac{V(u_h)}{V'(u_h)}+\frac{1}{6}\left(\frac{V(u_h)V''(u_h)}{V'(u_h)^2}-1   \right)(u-u_h)\\ \nonumber
&+& \mathcal{O}[(u-u_h)^2].
\label{series}
\end{eqnarray}
Note also that Eq.\ (\ref{constraints}) implies that 
\begin{equation}
T=\frac{e^{A(u_h)+B(u_h)}|V'(u_h)|}{4\pi} \,.
\label{hawking1}
\end{equation}
This equation will be used in our discussion about the Polyakov loop later on.

The main focus of this paper is to unveil some of the main properties displayed by Polyakov loops in deformed CFT's that have thermodynamic properties similar to 4 dimensional $SU(N_c)$ pure YM theories. The relevant operator for our discussion is the path-ordered Polyakov
loop defined as
\begin{equation}
\ell (\vec{x})=\frac{1}{N_c}{\rm Tr} \,P \,e^{i\int_{0}^{1/T} \hat{A}_{0}(\vec{x},\tau)d\tau}
\label{wilsonloop}
\end{equation}
$\hat{A}^{\mu}$ is the non-Abelian gauge field operator and the trace is over the fundamental representation of $SU(N_c)$. Since in YM theories the only degrees of freedom are adjoint gluons, no other adjoint scalar fields (which would describe the transverse excitations of the set of $N_c$ D3-branes) are included and we assume that the dual theory may be properly defined in a 5 dimensional spacetime. 

When $N_c \to \infty$ and the radius of the background spacetime is much larger than the string length, an infinitely massive excitation in the fundamental representation of $SU(N_c)$ in the gauge theory is dual to a classical string in the bulk hanging down from a probe D-brane at infinity \cite{Maldacena:1998im,Rey:1998bq,Bak:2007fk}. The endpoints of the strings behave as non-dynamical fundamental probes and they are the analog of the infinitely heavy quark limit in quenched lattice gauge theory. 

According to the gauge/string duality, the correlator of Polyakov loops is given by $\langle \ell (\vec{x})\ell^{*}(0)\rangle = Z_{string}$, where $Z_{string}$ is the full string generating functional that includes a sum over all the string worldsheets $\mathcal{D}$ (which may include connected and disconnected configurations) whose boundaries describe a $Q\bar{Q}$ pair separated by a distance $|\vec{x}|=L$ in the gauge theory \cite{Bak:2007fk}. This defines the $Q\bar{Q}$ heavy quark free energy via $e^{- \mathcal{F}_{Q\bar{Q}}(L,T)/T} = \langle \ell (\vec{x}) \ell^{*}(0)\rangle$. Within this approximation, the string dynamics (in Euclidean space) described by a given class of worldsheets $\mathcal{D}$ is given by the classical Nambu-Goto (NG) action 
\begin{equation}
\ S_{NG}(\mathcal{D})=T_s\int_{\mathcal{D}} d^2\sigma \sqrt{{\rm det} \,h_s^{ab}}
\label{nambugotoaction}
\end{equation}
where $T_s=\frac{1}{2\pi \alpha'}$ is the string tension, $h_s^{ab}=G_s^{\mu\nu}\partial^{a}X^{\mu}\partial^{b}X^{\nu}$
($a,b=1,2$), $X^{\mu}=X^{\mu}(\tau,\sigma)$ is the embedding of the string in
the 5-dimensional spacetime, and $G_s^{\mu\nu}$ is the background bulk metric in the string frame. In general, the metric in these frames are related as follows: $G_s^{\mu\nu}=e^{\xi \phi}G^{\mu\nu}$, where $\xi>0$ \cite{comment1}.    

The connected contribution to the correlator, $\langle \ell (\vec{x})\ell^{*}(0)\rangle -|\langle \ell \rangle|^2$, is $\sim \mathcal{O}(N_c^0)$ \cite{Bak:2007fk}. In $\mathcal{N}=4$ SYM, the connected configuration that minimizes the action when $LT \ll 1$ is a U-shaped curve that connects the string endpoints at the boundary and has a minimum at some $u^*$ in the radial coordinate \cite{Rey:1998bq}. The connected part of the correlator in models such as (\ref{dilatonaction}) was recently discussed in \cite{Alanen:2009ej}.

In a confining theory, below $T_c$ only connected worldsheets should be considered and, thus, in the supergravity limit the U-shaped configuration should be able to describe the full correlator at any $L$ and $T\leq T_c$ (other type of connected configurations become important above $T_c$ at large $L$ \cite{Bak:2007fk}). In these theories, it can be shown that in the limit where $L\to \infty$ the vacuum heavy quark potential becomes $\mathcal{V}_{Q\bar{Q}}(L)=\sigma_0 L +E_0 +\mathcal{O}(1/L^\delta)$, where $\delta >0$, $\sigma_0$ is the string tension, and the zero point energy $E_0 <0$ \cite{Kinar:1998vq} (the zero point energy in QCD was recently discussed in \cite{Hidaka:2009xh}). Roughly speaking, in the dual picture confinement occurs because the backreaction in the geometry given by the scalar prevents the bottom of the string to probe the deep IR region (as one pulls the quarks apart the only thing the dual string can do is to increase its length by the same amount). Thus, confinement implies that $|\langle \ell\rangle|=0$ when $T\leq T_c$. Similarly, the imaginary part of the Wilson loop induced by thermal fluctuations, which was recently derived in \cite{Noronha:2009da}, can be shown to be identically zero in confining theories when $T\leq T_c$. 

The disconnected piece of the correlator is the only $\mathcal{O}(N_c^2)$ contribution when $N_c$ is large. In the gravity picture, this corresponds to two infinitely distant and, hence, independent straight strings going from the boundary of the space down to the horizon where the string endpoints can attach to the different $N_c$ D3-branes, which shows that this contribution indeed goes as $N_c^2$. Note that the $|\vec{x}|\to \infty$ limit of the full correlator is described by two different classes of worlsheets depending on $T$: below $T_c$ one may use the connected U-shaped worldsheet ($\sim \mathcal{O}(N_c^0)$) while above $T_c$ cluster decomposition says that in this limit the full correlator equals the disconnected piece, which is $\sim \mathcal{O}(N_c^2)$. Therefore, in models such as (\ref{dilatonaction}) confinement implies that $|\langle \ell \rangle|$ jumps at the transition, which must be of first-order. Since the deconfinement phase transition in pure glue is expected to be first-order when $N_c \geq 3$, the class of gravity theories discussed here may indeed provide a description of the phase transition in YM theories at large $N_c$. 

We shall now show how to describe the disconnected contribution to the correlator in theories such as (\ref{dilatonaction}). Using the gauge choice $X^{\mu}=(t,x(u),0,0,u)$, the regularized free energy of a single (static) heavy quark (up to a $T$ independent constant) above $T_c$ is
\begin{equation}
\ F_{Q}(T) = T_s \int_{0}^{u_h}du\left(\sqrt{M(u)}-\sqrt{M_0(u)}\right)-T_s \,u_h^{\frac{1}{\Delta-4}}
\label{singlefenergy}
\end{equation}
where $M(u)\equiv |G_{s\,00}|G_{s,\,uu}$ is a positive-defined function of $u$ that is regular at the horizon. Near the boundary this function is $M_0(u)=\lim_{u\to 0}M(u)=u^{\frac{10-2\Delta}{\Delta-4}}/(4-\Delta)^2$ as long as $V$ obeys Eq.\ (\ref{conditionboudnary}). The Polyakov loop is given by 
\begin{equation}
\ |\langle \ell\rangle| = e^{-F_{Q}(T)/T} \,.
\label{ploop}
\end{equation}   
   
Instead of using Eq.\ (\ref{singlefenergy}), it is more convenient to work with its derivative $dF_{Q}/dT=\left(dF_{Q}/du_h\right) \,du_h/dT$. In fact, 
\begin{equation}
\frac{dF_{Q}}{du_h}= T_s \sqrt{M(u_h)} = T_s \,e^{\sqrt{\frac{2}{3}}\,u_h}\,e^{A(u_h)+B(u_h)} 
\label{dFduh}
\end{equation}
and one obtains that
\begin{equation}
\frac{dF_{Q}}{dT}=T_s\, e^{\sqrt{\frac{2}{3}}u_h}\, e^{A(u_h)+B(u_h)}\frac{du_h}{T d\ln T}\,.
\label{dFdT}
\end{equation}
The speed of sound in the plasma is $c_s^2=d \ln T/d \ln s $, which implies that $du_h/d\ln s= c_s^2 du_h/d\ln T$. Therefore, one can rewrite the equation above explicitly in terms of $c_s^2$
\begin{equation}
\frac{dF_{Q}}{dT}=T_s \,e^{\sqrt{\frac{2}{3}}u_h} \,e^{A(u_h)+B(u_h)}\frac{s}{T\,c_s^2}\frac{du_h}{ds}\,
\label{dFdT1}
\end{equation}
On the other hand, one can use Eq.\ (\ref{bekenstein}) to show that $ds/du_h=3A'(u_h)s=3 G(u_h) s$. This identity, together with the horizon condition in Eq.\ (\ref{constraint1}) and the formula for the temperature in (\ref{hawking1}), shows that in gravity duals governed by Eq.\ (\ref{dilatonaction}) the following equation holds
\begin{equation}
\frac{dF_{Q}}{dT}= 4\pi T_s \frac{e^{\sqrt{\frac{2}{3}}u_h(T)}}{V(u_h(T))}\frac{1}{c_s^2(T)} \,.
\label{dFdT1}
\end{equation}
This is the main result of this paper. It shows that $dF_{Q}/dT < 0$ (as expected from thermodynamics) and, consequently, that $|\langle \ell \rangle|$ is a monotonically increasing function of $T$. For any potential that satisfies (\ref{conditionboudnary}) one finds that in the small $u_h$ (high-temperature) limit $u_h=(\pi R T)^{\Delta-4} $. Thus, in the high temperature limit $F_{Q} = -\frac{R^2}{2\alpha'}T$ for any $V$ (apart from a temperature independent constant and subleading terms that vanish at high $T$). This linear dependence in $T$ is also observed in conformal theories \cite{Noronha:2009ia}, which agrees with the fact that the spacetime considered here is asymptotically identical to $AdS_5$.

Since $V(u_h)$ is well defined at finite $u_h$, one can see that $dF_{Q}/dT$ only diverges when $c_s \to 0$. Moreover, for confining gravity duals the magnitude of the jump in $|\langle \ell \rangle |$ at $T_c$ (also known as the Gross-Witten point \cite{adrianrob}) is related to how quickly $c_s$ vanishes at $T\to T_c^{+}$. Continuous transitions where $|\langle \ell \rangle|$ smoothly interpolates between zero and its high-T limit, which are expected to occur in the presence of dynamical fermions, can also be obtained depending on the choice for $V(\phi)$. 

The class of potentials (which satisfies the BF bound in the UV and also leads to an interesting behavior at finite temperature) relevant for YM we use here is of the form \cite{Gubser:2008ny} 
\begin{equation}
\ V(\phi)=\frac{-12\,\cosh \gamma \phi+b_1\phi^2}{R^2},
\label{potential}
\end{equation} 
where $\gamma$ is directly related to the speed of sound in the IR and $b_1$ determines $\Delta$ near the UV fixed point. The choice $\gamma=\sqrt{1/2}$ and $b_1=1.95$ gives $\Delta=3.3784$ and leads to a black hole solution with the thermodynamic properties shown in Fig.\ 1, which can mimic the lattice data for the pressure $p$ in $SU(3)$ pure glue above $T_c$ as computed by Boyd et al. \cite{Boyd:1996bx} above $T_c$. This is done by computing the entropy density using Eq.\ (\ref{bekenstein}) and choosing the value for $k_5^2=0.150376$, which leads to the best agreement with the data. The other thermodynamic quantities are obtained from the entropy density via the usual thermodynamic identities.    
 
The gauge theory for this choice of parameters does not exhibit confinement (when $\xi=\sqrt{2/3}$) but has a very sharp (though continuous) transition near $T_c$, which is determined by the minimum of $c_s^2$. The trace anomaly $\theta \equiv \varepsilon-3p$, where $\varepsilon$ is the energy density, displays an interesting feature above $T_c$ when divided by $(T\,T_c)^2$: it plateaus in the temperature range between $T/T_c=1-4$ where non-perturbative contributions to thermodynamic functions are expected to appear \cite{Pisarski:2006yk,megiasall}. At higher temperatures improved hard-thermal-loop perturbation theory is expected to provide a good description of the data \cite{Andersen:2009tc}. 

Because of our choice for $V$, the system is always in the deconfined phase and $|\langle \ell \rangle|$ can be computed at any $T$ using (\ref{dFdT1}) and there is no jump at $T_c$. The renormalized Polyakov loop in quenched $SU(3)$ QCD was computed on the lattice in \cite{Kaczmarek:2002mc} and more recently in \cite{Gupta:2007ax}. We compute $|\langle \ell (T) \rangle|$ by integrating Eq.\ (\ref{dFdT1}) and, because of the regularization, $F_Q$ is determined only up to a constant whose value is chosen via a match to the lattice data at $T=1.03\, T_c$ (the first data point above $T_c$ in Ref.\ \cite{Gupta:2007ax} for $N_t=4$) for a given choice of $\alpha'$. Our results for the renormalized Polyakov loop are shown in Fig.\ 2 (with $\alpha'/R^2=3.5$) and one can see they agree very well with the lattice calculations from \cite{Gupta:2007ax} between $T/T_c=1-3$. At higher temperatures the curve starts to considerably deviates from the data, which is expected since the gauge theory considered here is not asymptotically free.  

In the class of dual theories considered here $R^4/\alpha'^2$ is not exactly the t'Hooft coupling as it occurs in $\mathcal{N}=4$ SYM where the string dual is known. Here, its value could be fixed, for instance, by imposing that the ratio $T_c/\sqrt{\sigma_0}$ computed in a confining theory matches lattice results. However, for the non-confining theory discussed above $R^2/\alpha'$ is varied in order to find the optimal value that leads to the best description of the lattice data. Since this value turns out to be $\sim\mathcal{O}(1)$, one should expect that higher-order derivative corrections to the gravity action, such as those studied in \cite{higherderivative}, are important. Because of these corrections, $\eta/s$ will acquire an explicit $T$ dependence due to the coupling with $\phi$ evaluated at the horizon \cite{Cai:2009zv} (without these extra terms $\eta/s=1/(4\pi)$ regardless of $T$ \cite{KSS}). In QCD, $\eta/s$ is expected to have a minimum near $T_c$ \cite{csernai}. It would be interesting to find the minimal set of higher-order derivative corrections to (\ref{dilatonaction}) that can force this ratio to have a similar behavior in the dual gauge theory.   

\begin{figure}
\centering
\epsfig{file=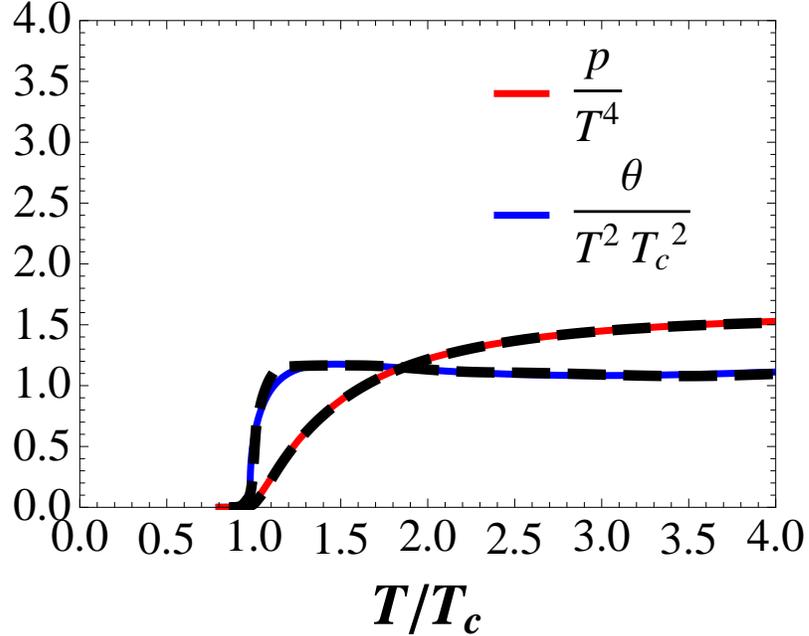,width=0.7\linewidth,clip=}
\caption{Comparison between the pressure and the trace anomaly of the black hole solution of Eqs.\ (\ref{dilatonaction}) and (\ref{potential}) (solid lines) and the lattice data (dashed lines) from Ref.\ \cite{Boyd:1996bx}. Only an interpolation of the lattice data is shown (the inclusion of the typical error bars would not lead to any significant changes in our analysis).}
\label{fig:SSB1}
\end{figure}

\begin{figure}
\centering
\epsfig{file=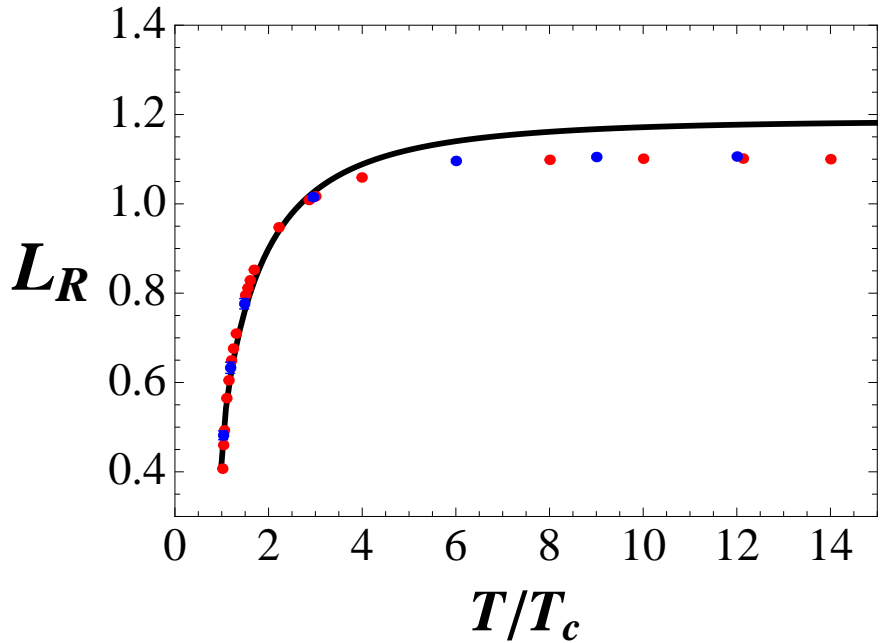,width=0.7\linewidth,clip=}
\caption{The renormalized Polyakov loop (solid black line) for the gravity dual in Eqs.\ (\ref{dilatonaction}) and (\ref{potential}) and the lattice data for $N_t=4$ (red) and $N_t=8$ (blue) from Ref.\ \cite{Gupta:2007ax}.}
\label{fig:SSB2}
\end{figure}


I thank M.~Gyulassy and A.~Dumitru for many helpful comments and encouragement. I also thank G.~Torrieri, S.~Gubser, A.~Yarom, A.~Nellore, P.~Petreczky, O.~Andreev, R.~Pisarski, W.~A.~Zajc, and M.~Panero for discussions. This work was supported by the US-DOE Nuclear Science Grant No.\ DE-FG02-93ER40764.



\begin{thebibliography}{99}
\bibitem{comment1} Ref.\ \cite{Gursoy:2007cb} considered the more ``standard" notation where the string frame action is $S=\frac{1}{2 k_5^2} \int d^5x\sqrt{-G_s}e^{-2\Phi}\left[\mathcal{R}_s+4(\partial \Phi)^2+\ldots\right]$, which shows that this action was computed at tree level perturbation theory and, as such, there is a prefactor of $1/g_s^2$, where $g_s=e^\Phi$ is the string coupling constant. Thus, in this case $G_s^{\mu\nu}=e^{4\Phi/3} G^{\mu\nu}$. In our notation, $\Phi=\sqrt{3/8}\,\phi$.
\end{thebibliography}
\end{document}